\documentclass[aps,pra,twocolumn,showpacs,superscriptaddress]{revtex4}
\usepackage{times,amsmath,amssymb,amsbsy,epsfig,color,graphicx,multirow}

\begin{document}

\title{Stochastic origin of Gompertzian growths}

\author{E. De Lauro}
\affiliation{Dipartimento di Matematica e Informatica,
Universit\`a degli Studi di Salerno, Via Ponte don Melillo,
I-84084 Fisciano (SA), Italy}

\author{S. De Martino}
\affiliation{Dipartimento di Matematica e Informatica,
Universit\`a degli Studi di Salerno, Via Ponte don Melillo,
I-84084 Fisciano (SA), Italy} \affiliation{INFN Sezione di Napoli, Gruppo collegato di Salerno,
Fisciano (SA), Italy}

\author{S. De Siena}
\affiliation{Dipartimento di Matematica e Informatica,
Universit\`a degli Studi di Salerno, Via Ponte don Melillo,
I-84084 Fisciano (SA), Italy} \affiliation{INFN Sezione di Napoli,
Gruppo collegato di Salerno, Fisciano (SA),
Italy}\affiliation{CNR-SPIN, Sede di Salerno, and CNISM, Unit\`a
di Salerno, Fisciano (SA), Italy}

\author{V. Giorno}
\affiliation{Dipartimento di Matematica e Informatica,
Universit\`a degli Studi di Salerno, Via Ponte don Melillo,
I-84084 Fisciano (SA), Italy}


\begin{abstract}
This work faces the problem of the origin of the logarithmic
character of the Gompertzian growth. We show that the macroscopic,
deterministic Gompertz equation describes the evolution from the
initial state to the final stationary value of the median of a
log-normally distributed, stochastic process.  Moreover, by
exploiting a stochastic variational principle, we account for
self-regulating feature of Gompertzian growths provided by
self-consistent feedback of relative density variations. This well
defined conceptual framework shows its usefulness by allowing a
reliable control of the growth by external actions.

\end{abstract}

\maketitle

\section{Introduction}
The Gompertz model, in its original conception, was born as
phenomenological one namely describing the observed age tables of
humans \cite{Gom}. In fact, B. Gompertz concluded his empirical
studies of tables introducing the distribution of human ages for a
given community, the now well known function

\begin{equation}
P(\tau)=\alpha e^{-e^{c-\beta \tau}}\label{eq1}
\end{equation}
where $\alpha > 0$, $\beta < 0$ and $c$ are constant.

It is interesting to note that Gompertz posed at the core of his
deduction the properties of geometrical progression; ``This law of
geometrical progression pervades, in an approximate degree, large
portions of different tables of mortality''. The relevance of the
geometrical progression in the framework of the natural phenomena
in a variety of experimental environments was pointed out at the
end of nineteenth century in the works of Galton \cite{Gal} and
McAlister \cite{McA}. They showed that the geometrical mean
(median) describes the behavior of a large set of natural
phenomena better than the arithmetic one. We note that Gompertz
law according to the Gompertz observation emerges from the
equilibrium between geometrical series associated to degradation
and an arithmetical progression ruling indefinite (Malthusian)
growth having the experimental observations to be made on suitable
time intervals. The characteristics of simplicity of Gompertz law
due to this general and profound mathematical framework, attract
the attention of nascent biological disciplines, where the growth
studies go back to the thirty years of last century. Since this
distribution has had so remarkable success in a variety of very
different situations that a lot of literature refers it simply as
``law of growth''.

The laws of growth of natural systems, and the deep origin of
their characteristic scales of, e. g., length, mass, energy, or
numerosity, are, now, intensively investigated in many branches of
science, such as biomedicine \cite{Koz,deV}, economy \cite{Fig},
population dynamics \cite{Mul}, astrophysics and cosmology
\cite{DeM}. So, starting from 1930 the Gompertz equation has
became one of the most used tool to account for mechanisms of
growth in a variety of systems in many fields
\cite{Bay,Baj,Bru,deV,Cow,Mul}. If we exclude, for example, from
incomes distribution the little percentage of rich people (1\%)
Gompertz law is the fitting distribution. Obviously so general
applicability has developed a very interesting debate regarding
the origin of its logarithmic structure. The arguments called down
to this aim look at the main aspects of the underlying systems
i.e., biological social or/and economics.

General theory of dynamical systems, quiescence, cell kinetics
theory, entropic and thermodynamical arguments have been advocated
and illustrated by many authors in a variety of interesting papers
along many years introducing also suitable generalizations and
connections with other growth models as logistic one. A good
synthetic description with a large bibliography can be considered
that of Bajzer et al (1997)\cite{Baj}.

We remark three main aspects of the delineated problem that we
consider relevant.

The first one; although, it can be significant to start attaching
the problem from specificity of a discipline, in fact this can
enlighten nodal points, the arguments bringing to logarithmic
behavior must be so general as well as is the applicability of
Gompertz law;

The second one; Gompertz curve cannot be other than a ``suitable''
description of a mean behavior of systems under studies that are
all characterized by a basic stochasticity.

The third one; many of the considered systems reach the limiting
size exploiting self-controlled evolutions. We take these
considerations as starting points to recovering the features, in
particular logarithmic behavior, of Gompertz equation. Before to
go into the details of our deduction it is usefulness to make some
preliminary consideration and as first step to give a description
of deterministic Gompertz equation also to establish our notation.

\section{Gompertz equation}

The standard form of the deterministic Gompertz equation is:

\begin{equation}
z^{-1}\frac{d z}{dt} = \beta - \alpha \ln{(\frac{z}{{\tilde z}}}),
\label{Gompertzeq1}
\end{equation}

where $z$ describes the ``size'' of some quantity characterizing
the system, $\beta$ and $\alpha$ denote two positive constants
with the dimensions of the inverse of time, and ${\tilde z}$ is a
constant which the same dimensions of $z$. \\
Eq. (\ref{Gompertzeq1}) can be recast as:
\begin{equation}
\frac{d (\ln{s})}{dt} = - \alpha \ln{s}, \label{Gompertzeq2}
\end{equation}
where $s(t) \doteq z(t)/z_{\infty}$, and $z_{\infty} = {\tilde z}
\exp{(\beta/\alpha)}$. The Gompertz equation is then associated to
four parameters (all dependent on the specific system): $\alpha$,
$\beta$, ${\tilde z}$, and the initial condition (``scale'') $z(0)
= z_0$. Its solution is:
\begin{equation}
z(t) = z_{\infty} \exp{[(\ln\gamma) \cdot e^{-\alpha t}]},
\label{Gompertzsol}
\end{equation} where $\gamma \doteq z_0/z_\infty$. It is immediately verified
that this solution always approaches monotonically in time
$z_\infty$: depending on the conditions $z_0 < z_\infty \equiv
{\tilde z} \exp{\beta/\alpha} \; \; \; (\gamma <1)$, or $z_0 >
z_\infty \equiv {\tilde z} \exp{\beta/\alpha} \; \; (\gamma
> 1)$, the system  monotonically grows or monotonically decreases,
respectively, from the dimension $z_0$ to the dimension
$z_\infty$, approaching the asymptotic value $z_\infty$ with the
characteristic time $\alpha^{-1}$. It is worth noticing that the
solutions of the eq. (\ref{Gompertzeq2}) satisfy some, somewhat
simple and self evident properties, deriving by peculiar features
of logarithmic function. These properties justify why the Gompertz
equation plays a major role among all the equations describing
growth phenomena.

If we characterize a solution $s(t)$ by the pair of values of its
parameters $(\gamma, \alpha)$, the properties are the following:

\begin{enumerate}
    \item the product $s_1 (t) \cdot s_2 (t)$ of two solutions with
parameters $(\gamma_1, \alpha)$ and $(\gamma_2, \alpha)$,
respectively, is again a solution with parameters $(\gamma_1 \cdot
\gamma_2, \alpha)$,

    \item if $s(t)$ is a solution with parameters $(\gamma, \alpha)$,
then $s^a (t)$, with $a \in {\bf R}$ is a solution as well, with
parameters $(\gamma^a, \alpha)$.

    \item the constant function $s(t) = 1$ is a (trivial) solution.
\end{enumerate}

Note that, for $a =-1$ we obtain that the inverse of $s(t)$,
$s^{-1}(t)$ is a solution associated to $(\gamma^{-1}, \alpha)$.
Then it is straightforward to verify that if for example the
original solution is obtained with $z_0 > z_\infty$ (aggregation
process) $s^{-1}(t)$ describes the time-reversed (fragmentation)
process from $z_\infty$ to $z_0$.  Note also that this property,
together with properties 1), 3) makes the set of solutions of the
Gompertz equation an Abelian group. Note finally that the property
3) implies that any other quantity $M(t)$ linked to $z(t)$ by an
allometric relation $(M(t)=b \cdot z^a(t))$ satisfies the same
Gompertz equation with modified parameter $\gamma^a$

\section{Gompertz equation as evolution equation for the median of geometrical Brownian motion.}

\subsection{Lognormality as basic in a variety of natural
systems}

In 1947 H.R. Jones linked the problem of mortality to life
expectancy and ageing processes \cite{Jon}. His germinal point of
view was that diseases and disfunctions accumulate slowly along
the time damaging \emph{multiplicatively} the human bodies. The
extensive analysis of Jones showed that Gompertz equation applies
exactly to people that have not eliminated the first cause of
diseases, i.e. hygienic condition. A good mean improvement of
these last one, namely slight, modifies the general behavior.
Then, following Jones, at the basis of the analysis of life
expectancy, there is a stochastic process built with independent
random variables (diseases and/or social and economic condition)
that add multiplicatively. Moreover at the end of sixties years of
the last century a detailed statistical analysis performed by
Sachs showed that physiological parameters like blood pressure,
tolerability of medicaments, body size survival rate are
lognormally distributed \cite{Sac}. Finally it is worthwhile to
note that the lognormal i.e. geometric Brownian motion appears to
be asymptotic to a variety of branching processes introduced  to
describe cell systems growth \cite{Dei,Cow}.

 On the other hand it is well known that lognormal distribution arises in a
variety of classification procedures and in physical and
biological systems when natural genesis involves repeated
breakages or aggregations. Very relevant, in 1941, Kolmogorov
\cite{Kol} has shown that when the frequency of
aggregation-disaggregation in a growth process is independent of
the size of the constituents, the asymptotic size distribution of
the aggregate should tend to be lognormal. This so general
multiplicative behavior of basic randomness of underlying
processes has  been indicated  as a ``Multiplicative
Gestaltungs-Principle of Nature''. It is also assumed as its
possible accounting a general property
 of ``coherence''  of natural  systems \cite{Pop}.
The simple considerations outlined above, at the light of the two
first points of our introductory section namely basic
stochasticity of systems and large applicability of the Gompertz
model, bring us to postulate that the Gompertz equation is the
deterministic one emerging from geometric Brownian motion, that is
the general stochastic structure associated with a variety of
systems. The necessary step at this point is to give a simple
mathematical procedure connecting stochastic structure with
deterministic one.

\subsection{Mathematical procedure}

The main feature of lognormal can be so summarized: a random
variable X is said to have a lognormal distribution with suitable
parameters associated to mean and variance if $\ln\{(X(t))\}$ is
normally distributed. Consider then the diffusion process $X(t)$
taking non negative values, and satisfying the Ito differential
equation:

\begin{equation}
dX(t) = \{\lfloor\frac{\nu}{2}X(t)-\alpha
X(t)\ln{\bigl[\frac{X(t)}{K}\bigr]}\}dt + \sqrt{\nu} X(t) dW(t),
\label{Gompertzdiffproc}
\end{equation}where $\alpha$ and $\nu$ are positive constants, $K$ has the same
dimensions of $X(t)$, and $dW(t)$ denotes a Gaussian stochastic
process (Wiener process) with zero mean and variance $dt$. Here
$X(t)$ can describe any quantity. X(t) is a multiplicativy
diffusive process (geometrical brownian motion) and its Ito
equation can be recast as:
\begin{equation}
dX(t) = -\alpha X(t)\ln{\bigl[\frac{X(t)}{K}\bigr]} dt +
\sqrt{\nu} X(t) dW(t), \label{Gompertzdiffproc}
\end{equation} where $$K=\tilde{K}e^{\frac{\nu}{2\alpha}}$$.

We now prove that this process generates the deterministic
Gompertz equation. Let us in fact associate to $X(t)$ a new
process $Y(t)$ defined by the relation

\begin{equation}
\frac{X(t)}{K} \equiv e^{Y(t)}, \label{AssOUprocess}
\end{equation} where $Y(t)$ is adimentional and takes values ranging on
($-\infty, +\infty$) when X(t) takes values on [$0, \infty$ ). By
exploiting Eq. (\ref{Gompertzdiffproc}) and definition
\ref{AssOUprocess}, we can compute $d (\exp{Y(t)})$: by Ito's
lemma obtaining:
\begin{equation}
dY(t) = -\alpha Y(t) dt + \sqrt{\nu} dW(t), \label{AssOUprocessEq}
\end{equation} We see that $Y(t)$ is an {\it Ornstein-Uhlenbeck process}. Its
probability density $p(y, t; y_0, 0) \equiv p(y, t)$, satisfies
the Fokker-Plank equation
\begin{equation}
\partial_t p(y, t) = \alpha \partial_y [y p(y, t)] +
\frac{1}{2} \nu_r \partial^2_y p(y, t), \label{FPOU}
\end{equation}
and can be exactly computed for any initial condition \cite{Gar}:

\begin{equation}
\int_{-\infty}^{-\infty} du e^{-i u y}\chi_0 (u e^{-\alpha t})
e^{\bigl[-\frac{\nu_r u^2}{4 \alpha}(1-e^{-\alpha} t)\bigr]},
\label{ProbDensOU}
\end{equation} where $\chi_0 (u)$ is the characteristic function of the initial
probability. The moments $<Y^n (t)>$ of the process $Y(t)$:

\begin{equation}
<Y^n (t)> \; \; \doteq \; \int_{-\infty}^{-\infty} dy \; y^n \;
p(y, t), \label{MomentsDef}
\end{equation} satisfy the set of {\it branching equations}

\begin{equation}
\frac{d}{dt} <Y^n (t)> = - n \alpha <Y^n (t)> + \frac{1}{2}
\sigma_r n (n-1) \alpha <Y^{n-2} (t)>, \label{MomentsEqs}
\end{equation}
which can be solved, iteratively at any finite order. We now focus
on the case $n=1$:
\begin{equation}
\frac{d}{dt} <Y (t)> = - \alpha <Y (t)>, \label{FirstMoment}
\end{equation}
exploit the relation (\ref{AssOUprocess}), and obtain
\begin{equation}
\frac{d}{dt} <ln \bigl[\frac{X(t)}{K}\bigr]> = - \alpha <ln
\bigl[\frac{X(t)}{K}\bigr]>. \label{PreGompertz}
\end{equation}
If we define the time-dependent quantity $z(t)$ by the relation
\begin{equation}
\frac{z(t)}{K} = \exp{^{<ln \bigl[\frac{X(t)}{K}\bigr]>}},
\label{PreGompertz}
\end{equation}
we see that $z(t)$, which has the same dimensions of $X(t)$,
satisfies the Gompertz equation, where the constant $K$ is
identified with $z_{\infty}$, i. e. with the asymptotic value of
$z(t)$. But {\it what about the meaning of $z(t)$?}. We nown
provide a precise meaning  to $z(t)$. Let us suppose that the
initial probability of the process $Y(t)$ is Gaussian (for
example, the solution is the fundamental one with an initial delta
function condition). By Eq. (\ref{ProbDensOU}), the solution at
any time is Gaussian:
\begin{equation}
p(y, t) dy = \frac{1}{\sqrt{2 \pi \sigma(t)}} e^{-\frac{1}{2
\sigma(t)}(y - \mu (t))^2} dy, \label{GaussianYSol}
\end{equation}
where $\sigma (t) \doteq <[Y(t) - \mu (t)]^2>$, and $\mu (t)
\doteq <Y (t)>$. By exploiting the relation (\ref{AssOUprocess}),
we can recast Eq. (\ref{GaussianYSol}), obtaining the probability
of the process $X(t)$ as
\begin{equation}
p(x, t) dx = \frac{1}{\sqrt{2 \pi \sigma(t)}} e^{\bigl[-\frac{1}{2
\sigma(t)}\bigl(\ln{\bigl(\frac{x}{K}\bigr)} - \mu
(t)\bigr)^2\bigr]} \frac{dx}{x}. \label{GaussianYSol}
\end{equation}
where, by solving the first two branching equation
(\ref{MomentsEqs})), mean $\mu(t)$
 and variance $\sigma (t)$ are given by:

\begin{equation}
\mu(t) = \mu_0 e^{-\alpha t},
\end{equation}

\begin{equation}
\sigma(t) = \sigma_0 e^{- 2 \alpha t} + \frac{\nu}{2 \alpha} (1 -
e^{- 2 \alpha t}). \label{Vart}
\end{equation}

We see that the process $X(t)/K$ is lognormally distributed. But
it is well known that in a lognormally distributed process the
mean of the logarithm of the process is the logarithm of the {\it
median} of the process. Then, we can conclude that if the process
$Y(t)$ is initially Gaussian, i.e. if the process $x(t)/K$ is
initially lognormally distributed, this last process remains
lognormally distributed at any time, and the variable $z(t)/K$
(which we denoted $z(t)/z_{\infty}$) {\it is the median of this
process}.

In conclusion we have proved that: \emph{the deterministic
Gompertz equation is the macroscopic consequence of a lognormally
distributed, diffusion process X(t); the macroscopical size z(t)
whose evolution is ruled by the Gompertz equation is the median of
process $X(t)$.}

It follows that the multiplicative stochastic process (a standard
geometric brownian motion) that many times is introduced as
``Gompertz stochastic tumor growth model'' is not due to an extra
noise disturbing the Gompertz growth, but it is itself the origin
of the deterministic growth.

To further support this conclusion we remark that deterministic
Gompertz model was extracted by B. Gompertz, just looking to the
properties of a geometric progression that emerges from the
mortality tables; i.e. the Gompertz function has its natural
interpretation as median of a multiplicative process.

\subsection{Short remark on the observability of microscopic parameters}

The underlying process $X(t)$ is characterized by three
parameters: $\alpha, \nu, K$. Their determination completely
define the process. As a first observation, we note that the
Malthusian parameter $\beta$ in the standard form
(\ref{Gompertzeq1}) of the Gompertz equation is provided, in a
consistent way, by the diffusion parameter: $\beta = \nu/2$. We
remark also that the drift parameter $\alpha$ is  clearly
macroscopically observable by fitting the macroscopic size growth
of the selected systems. Therefore, if one is able to provide a
method to extract by observational data the diffusion parameter
$\nu$, the whole process $X(t)$ can be reconstructed. To this end,
we consider that the observed size (i. e., the median of the
process) must displace a range of variability; in fact

\begin{equation}
\ln z(t)/z_{\infty} \equiv \ln z(t)/K \equiv <\ln X(t)/K> \equiv
<Y(t)>
\end{equation} \noindent is the mean of a Gaussian process with
variance (width) $\sigma (t)$. And, in particular, when the last
stage $K \equiv z_{\infty}$ is reached, the width (i.e variability
of the size) becomes that of the stationary distribution (see Eq.
(\ref{Vart})). If one performs a statistics of the observed sizes,
one could find that the sizes of a system at the last stage range
from a minimum one $z_{min}$ to a maximum one $z_{max}$, with the
following relations:
\begin{equation}
z_{\infty} = \sqrt{z_{min} \cdot z_{max}},
\end{equation}
(just geometric mean),
\begin{equation}
z_{min} = \frac{z_{\infty}}{r}, \; \; z_{max} = z_{\infty} \cdot
r, \; \; \\
r = e^{\frac{\nu}{2 \alpha}},
\end{equation} \noindent and
 \begin{equation}
 \nu = \alpha \ln{\frac{z_{max}}{z_{min}}}.
\end{equation} \noindent Summing up: $\alpha, z_{max}, z_{min}$ can
be extracted by observational data, and, in principle, their
values provide an estimate of all the other parameters of the
underlying process; in particular, of the order of magnitude of
the diffusion parameter $\nu$.

\section{Characterizing the Gompertzian growth process}

The conclusions reached in the previous section ascribe the
Gompertz equation within the framework of ``Multiplicative
Gestaltung-Principle of Nature'', i.e. to the ubiquitous nature of
lognormal distribution, which characterizes a variety of natural
system, i.e. those in which emerge characteristic scales of
``coherence''. The ``actual'' Gompertz equation is then the
stochastic one. The stochastic description however does not
exhaust all the aspects of Gompertzian model, namely stochasticity
allows spontaneous growth until a final size if and when systems
undergo to a stationary state. The Gompertzian growth is
characterized, on the contrary, as remarked in the third point of
introductory section, by a self-controlled  evolution ruled by
variation of density. Now we incorporate in our description this
relevant behavior of growing systems.

\subsection{Dynamical updating of geometrical Brownian motion}

In standard treatment the drift terms in a Fokker-Planck equation
is a function given a priori; backward and forward evolutions are
described with two different equations. The systems spontaneously
approach a stationary state when balancing is reached between the
stochastic term (Wiener process) and deterministic (drift);
consequently there is not, for example, time reversal invariance.
In order to have self-control (feedback) we must to add to
Fokker-Planck equation a dynamical equation updating the drift and
allowing time reversal behavior. The equation is the stochastic
equivalent of $\textbf{F}=m\textbf{a}$, and it can be written also
as the equation describing a suitable interface \cite{Row}.

We remark, that, being our systems self-controlled the dynamical
update must contain self coupling non linearity. We use a
stochastic variational principle theory. We recall this theory
briefly: deterministic dynamic evolutions are characterized by two
independent principles, the first kinematic the second dynamic.
The kinematic principle is provided by standard differential
rules, and the dynamic one by a variational principle, i.e. the
Lagrangian principle. The theory of stochastic variational
principle assumes Ito's equation as a kinematic rule and the
Lagrangian variational principle as dynamic rule. The variation is
made by considering conditional expectation. As a consequence: the
configuration of our systems is described, in general, by a
vectorial Markov process $\xi(t)$ taking values in $\Re^{3}$. This
process is characterized by a probability density $\rho ({\bf
r},t)$ and a transition probability density $p({\bf r},t|\,{\bf
r'},t')$, and its components satisfy an It\^o stochastic
differential equation of the form
\begin{equation}
d\xi_j(t)=v_{(+)j}\bigl(\xi(t),t\bigr)dt+d\eta_j(t),
\end{equation}
where $v_{(+)j}$ are the components of the forward velocity field.
As already observed here the fields $v_{(+)j}$ must not be given a
priori, but play the role of dynamical variables and are
consequently determined by imposing a specific dynamics. The noise
$\eta(t)$ is a standard Wiener process, $D$ is the diffusion
coefficient. We indicate by $E_{t}$ the conditional expectations
with respect to $\xi(t)$. In what follows, for sake of notational
simplicity, we will limit ourselves to the case of one dimensional
trajectories, but the results that will be obtained can be
immediately generalized to any number of dimensions. We will
suppose for the time being that the forces will be defined by
means of purely configurational, possibly time-dependent $V(x,t)$,
potentials, this includes also a non linear potential functional
of the density of the process. A suitable definition of the
Lagrangian and of the stochastic action functional for the system
described by the dynamical variables $\rho$ and $v_{(+)}$ allows
to select, the processes which reproduce the correct dynamics
\cite{Nel,Gue,Cuf}. In fact, while the probability density $\rho
(x,t)$ satisfies, as usual, the forward Fokker-Planck equation
associated to the stochastic differential equation
\begin{equation}
\partial_t \rho = D
\partial_x^2 \rho -\partial_x ( v_{(+)} \rho ) = \partial_x ( D
\partial_x \rho - v_{(+)} \rho )\label{FP}
\end{equation}

the following choice for the Lagrangian field

\begin{equation}
L(x,t)={m\over2}v_{(+)}^2 (x,t) + m D \partial_x v_{(+)}(x,t) -
V(x,t)
\end{equation}
enables to define a stochastic action functional
\begin{equation}
\mathcal{A}=\int_{t_{0}}^{t_{1}} E_{t} [ L( \xi(t),t)]dt
\end{equation}
which leads, through the stationarity condition $\delta
{\mathcal{A}}=0$, to the equation
\begin{equation}
\partial_tS +
{(\partial_xS)^2\over2m} + V \, - 2mD^2 \,
   {\partial^2_x\sqrt{\rho } \over \sqrt{\rho }} = 0 \, .\label{Madelung}
\end{equation}

The field $S(x,t)$ is defined as
\begin{eqnarray}
S(x,t)=-\int_t^{t_1}E\left[ L\big(\xi(s),s\big)\,\big|\,\xi(t)=x
\right]ds+ \nonumber \\ \,+E\left[ S_1\big(\xi(t_1)\big)\,\big |
\,\xi(t)=x \right]
\end{eqnarray}

where $S_1(\,\cdot\,)=S(\,\cdot\,,t_1)$ is an arbitrary final
condition. This equation is the well know equation of interfaces
theory where the term depending on density $ \rho $ represents the
contribute due to surface tension.

By introducing the function $R(x,t)\equiv \sqrt{\rho(x,t)}$ and
the de Broglie ansatz
\begin{equation}
\psi(x,t)=R(x,t)\,{\rm e}^{iS(x,t)/2mD}\label{DeBroglie}
\end{equation}
equation (\ref{Madelung})  takes the form
\begin{equation}
\partial_tS+{(\partial_xS)^2\over2m}+V\,-\,{2mD^{2}}\,
        {\partial^2_xR\over R}=0 \label{Madelung1}
\end{equation} and the pair of real equations (\ref{FP}) and (\ref{Madelung1})
are equivalent to the single linear  equation for  $\psi$

\begin{equation}
i(2mD)\partial_t\psi= -\,{2mD^{2}}\,\partial_x^2 \psi
   + V \psi,
\label{SL}
\end{equation}

This connects our dynamic  equations with an ordinary eigenvalue
problem on Hilbert space.

Note that the observables are the density $\rho$, which here
represents just the (mass) density of the system, and the drift
mean velocity of the system $v$; where the connection with the
pair $\psi, v$ is provided, at every point and at every time, by
$\rho = |\psi|^2$, and $v = \partial_x S/m$. Note also that
\begin{equation}
mv_{(+)}=\partial_x S + m D \,{\partial_x R\over R}.
\end{equation}
\noindent If we choose the potential in the form of
$V(x)=f(\rho)$, where $f(\rho)$) is a nonlinear functional of
density (e.g. $ln(\rho)$), we obtain a dynamic system that
provides self-regulation by the self consistent feedback action of
relative density variations typically associated to osmotic
phenomena across interfaces. So well this very general model
contains the third of relevant aspects considered to describe
growing systems, in particular cells proliferation in solid tumor.
Lognormal stochastic background, self-control and interface theory
are the constitutive elements of this model, which leads to the
deterministic Gompertz equation as equation for the median. This
approach, being conceptually well-founded, allows to face, in
principle, the growth problems as a controlled one namely it is
possible to intervene with outdoor control. For example our method
poses the reduction of tumoral mass from a new and interesting
point of view  as we describe in the next section.

\subsection{Controlled growth}

In this section we move on to implement the controlled evolution.
In fact we exploit the transition probabilities of the Gompertz
self-controlled  evolution to model controlled evolutions  from a
given initial state to arbitrarily assigned final states. We start
by observing that to every solution $\psi(x, t)$ of the
Fokker-Planck equation (\ref{FP}), with a given $v_(+) (x, t)$ and
constant diffusion coefficient $D$, we can always associate a
"wavefunction" descriptions of eq.(\ref{SL}), of a dynamical
system. To this aim, it is sufficient to introduce a suitable
time-dependent potential $V_c (x, t)$, by exploiting the wave
equation (\ref{SL}) as a control equation\cite{Cuf}.

Here we quickly recall those elements of the controlling procedure
which are needed for our aim, referring to Cufaro et al. 1999
\cite{Cuf} for further details. Let us consider a solution $\rho
(x,t)$ of the Fokker--Planck equation, with a given $v_{(+)}(x,t)$
and a constant diffusion coefficient $D$; let us introduce the
functions $R(x,t)$ and $W(x,t)$ defined by
\begin{equation}
\rho (x,t)=R^2(x,t)\,,\qquad\quad v_{(+)}(x,t) =\partial_x W(x,t),
\label{DeBroglie}
\end{equation}
and remind that the relation
\begin{eqnarray}
mv_{(+)}=\partial_x S + m D \,{\partial_x R\over R} \nonumber \\
\, \equiv
\partial_x S + {m D/2}\,{\partial_x \rho \over \rho } =
\partial_x\left(
S + {m D/2}\ln\tilde \rho \right)\label{eq35}
\end{eqnarray}
\noindent must hold, where $\tilde \rho$ is an adimensional
function (argument of a logarithm) obtained from the probability
density $\rho$ by means of a suitable and arbitrary dimensional
multiplicative constant. If we now impose that the function
$S(x,t)$ must be the phase of a wave function, we immediately
obtain  from the Eqs. (\ref{DeBroglie}) and (\ref{eq35}).

\begin{equation}
S(x,t)=m W(x,t)-\frac{m D}{2}\ln\tilde-\theta (t), \label{SfromRo}
\end{equation}
which allows to determine $S$ from $\rho$ and $v_{(+)}$ (namely
$W$) up to an additive arbitrary function of time $\theta(t)$.
However, we must ensure that the wave function (\ref{DeBroglie})
with $R$ and $S$ given above, is a solution of the wave equation
(\ref{SL}). Since $S$ and $R$ are now fixed, equation (\ref{SL})
must be considered as a relation (constraint) defining the
controlling potential $V_{c}$, which, after straightforward
calculations, yields
\begin{eqnarray}
&&V_{c}(x,t)={m D^2}\,\partial_x^2\ln \tilde \rho +
{m D}(\partial_t\ln {\tilde \rho} \nonumber \\
&&+ v_{(+)} \partial_x \ln {\tilde \rho)
-{mv_{(+)}^2\over2}}-m\partial_t W + \dot\theta.
\label{ControlPotGen}
\end{eqnarray}
Of course, if we start with a wave function $\psi(x,t)$ associated
to a given time--independent potential $V(x)$, the
self-consistency is ensured, and this formula always yields back
the given potential, as it should.

We now leave the general way, and focus on the very interesting
case, useful for our goal, of simple controlling potentials able
to produce a evolution which can vary, and in particular reverse,
the growth trend. We start by observing that, to our purposes, it
is expedient to handle the Gaussian Ornstein-Uhlenbeck process
$Y(t)$, Eq. (\ref{AssOUprocessEq}), because this is a simple task,
and  it is equivalent to modify the underlying non Gaussian
process $X(t)$. In fact, being $<Y (t)> = \mu (t) =
<\ln{(X(t)/K)}> =  ln z(t)$, where $z(t)$ is the median describing
the macroscopic size, the monotonicity of the logarithmic function
ensures that a reduction of $\mu(t)$ leads to a corresponding
(multiplicative) reduction of the size. The probability
distribution for $Y(t)$ is the Gaussian (\ref{GaussianYSol}),
characterized by the two time-dependent parameters ($\mu (t),
\sigma (t)$). Let us suppose now that the time evolution at some
instant ${\bar t}$ has led to the pair of values $\mu ({\bar t})
\equiv \mu_0, \sigma ({\bar t}) \equiv \sigma_0$, and that we aim
to reduce this values in such a way that, after some
characteristic time $\tau$, they becomes $\mu_1 < \mu_0, \;
\sigma_1 < \sigma_0$. As proved in Cufaro et al.\cite{Cuf}, in
this Gaussian instance the controlling potential to be applied is
harmonic, and has the form:
\begin{equation}
V_{c}(x,t)={m\over 2}\, \left[\omega^2(t)x^2-2a(t)x +c(t)\right].
\label{ControlPotGauss}
\end{equation}
This harmonic potential is completely determined by the choice of
mean and variance, $\mu_{c}(t), \sigma_{c}(t)$ , of the Gaussian
process self consistently generated by the controlling harmonic
potential (\ref{ControlPotGauss}). In fact, the time-dependent
coefficients $\omega^2(t), a(t), c(t)$ are all functions of
$\mu_{c}(t), \sigma_{c}(t)$ and of their time derivatives, but a
further free function of time that can be exploited to simplify
the expression of the potential or of the phase. Being we here
merely interested to outline the general conceptual frame allowing
control, we avoid burdening this subsection with the explicit
expressions of the time-dependent coefficients, which can be found
in Cufaro et al. \cite{Cuf}. We only sum up again the procedure.
One chooses the form of the controlling mean and variance
 $\mu_{c}(t), \sigma_{c}(t)$, which in the characteristic time
$\tau$ go to the final reduced values $\mu_1, \sigma_1$; then one
inserts their expressions, and a suitably chosen form of the
further free function of time, in the time dependent parameters
$\omega^2(t), a(t), c(t)$, so obtaining an harmonic controlling
potential that, applied to the system, drives it towards the
reduced mean and variance. We comment here on an important aspect:
one can choose many functions $\mu_{c}(t), \sigma_{c}(t)$ leading
to the reduced mean and variance. If we set $\mu_{c}(t) = \mu_0
f(t)$, and $\sigma_{c}(t) = \sigma_0 g(t)$, the functions $f(t),
g(t)$ must satisfy the constraints:

$$f(0) = 1, \; f(\infty) = \mu_0/\mu_1 \; \; ; \; \; g(0) = 1, \; g(\infty) = \sigma_0/\sigma_1.$$

A possible form for $g(t)$, for example, could be:

$$\frac{1 + b \exp{(t/\tau})}{c + d \exp{(t/\tau})},$$

\noindent with $c + d = 1 + b; \; \; d = b \cdot
(\sigma_0/\sigma_1)$. Another possible choice can be found in
\cite{Cuf}.  The important point is that the choice of this
functions, together with that of the further free function of
time, must be made in such a way to define a controlling potential
$V_c (x, t)$ which can be effectively engineered and applied to
the system. It is also evident that, from the point of view of
practical implementations, one can resort to suitable
approximations which can anyhow realize the goal within a
permissible error.

\section{Conclusion}

The ubiquity of the Gompertz equation, raises a very interesting
question in the framework of growth phenomena, namely the key
presence of the logarithmic function ruling the nonlinear growth.
In this work, we have shown that the ubiquity of this equation and
its logarithmic regulation are the macroscopic expression of the
ubiquity of the log-normal distribution. In fact, we have proved
that Gompertzian growths are generated by a log-normally
distributed stochastic process, being the macroscopic Gompertz
equation the evolution in time of the median of the process. The
median then describes the macroscopic size of the growing system.
We remark, therefore, that the growth is not, as often supposed,
{\it disturbed} by a stochastic noise, but that the stochastic
process is its {\it origin}, and its {\it guidance} at any time.
This scheme agrees with the claim of Galton \cite{Gal} and
McAlister \cite{McA} who at the end of the nineteenth century have
shown that many natural systems are well described by
log-normality; their actual behavior is then described by the
median rather than the mean, thus implying that the basic
geometric series  plays a key role in describing relevant natural
phenomena as already inferred by Gompertz \cite{Gom}. In other
words, the root causes do not add but multiply among them in many
systems, including economic and social ones. Our analysis accounts
for the stochastic origin of ubiquity of Gompertzian growths,
suggesting also a method to determine the order of magnitude of
the microscopic stochastic parameters, and in particular of the
diffusion parameter, by the statistical observation of macroscopic
sizes.

Some systems, as for example cell aggregates, develop by a
"birth-death" process, i. e. more general by the competition
between aggregation and disaggregation. For this system, the basic
process is then a branching process leading in a suitable limit to
a log-normal behavior. Some of these aggregates are characterized
by self-regulation: an example is that of tumor cells forced to
grow in three dimensions, which cannot go beyond a critical
diameter regardless of how often new medium is provided or how
much open space is made available. These growths cannot thus be
described by a stochastic background where the parameters are a
priori defined and one must resort to a dynamical setting. In the
last section of this work we then propose a dynamical conceptual
framework that include self regulation mechanisms. We exploit a
stochastic variational principle, whose canonical structure
generates a control equation ruling the dynamical update of the
forward velocity. Self regulation is then implemented by choosing
the potential function as a (nonlinear) functional of the density
of the system, so introducing a self consistent feedback of the
relative density variations. This conceptual framework opens the
way to a reliable control of the growing system by external
actions which, for example, suitably modify, on some
characteristic time scales, the density profile of the system. One
can thus effectively develop practical methods to reverse the
growing trend, reducing, for example, a cancer size. The general
scheme here outlined will be applied and tested in forthcoming
papers, focusing on the behavior of specific systems.




\end{document}